\documentclass{emulateapj}
\usepackage{apjfonts}

\usepackage{graphics}

\newcommand{\be}{\begin{equation}}
\newcommand{\ee}{\end{equation}}
\newcommand{\bd}{\begin{displaymath}}
\newcommand{\ed}{\end{displaymath}}

\shorttitle{On the dust tori in Palomar-Green quasars }
\shortauthors{Cao}

\begin{document}


\title{On the dust tori in Palomar-Green quasars}

\author{Xinwu Cao}
\affil{Shanghai Astronomical Observatory, Chinese Academy of
Sciences, 80 Nandan Road, Shanghai, 200030, China\\Email:
cxw@shao.ac.cn\\(Received 2004 July 4; accepted 2004 August 20)}

\begin{abstract}
The dust clouds in the torus of the quasar are irradiated by the
central source, and the clouds at the inner radius of the torus
re-radiate mostly in the near-infrared (NIR) wavebands. The ratio
of the near-infrared luminosity to the bolometric luminosity
$L_{\rm NIR}/L_{\rm bol}$ can therefore reflect the torus geometry
to some extent. We find a significant correlation between the
ratio of the near-infrared luminosity to the bolometric luminosity
$L_{\rm NIR}/L_{\rm bol}$ and the central black hole mass $M_{\rm
bh}$ for the  Palomar-Green(PG) quasars, whereas no correlation is
found between the Eddington ratio $L_{\rm bol}/L_{\rm Edd}$ and
the ratio $L_{\rm NIR}/L_{\rm bol}$. Similar correlations are
found for the mid-infrared and far-infrared cases. It may imply
that the torus geometry, i.e., the solid angle subtended by the
dust torus as seen from the central source, does not evolve with
the accretion rate. The correlation of the solid angle subtended
by the torus with the central black hole mass $M_{\rm bh}$ implies
that the formation of the dust torus is likely regulated by the
central black hole mass. We find that the torus thickness $H$
increases with quasar bolometric luminosities, which is different
with the constant torus thickness $H$ with luminosity assumed in
the receding torus model. The average relative thickness $H/R$ of
the tori in the PG quasars derived from the ratios of the infrared
to bolometric luminosities is $\sim 0.9$. The further IR
observations on a larger quasar sample including more fainter
quasars by the Spitzer Space Telescope will help understand the
physics of the dust tori in quasars.

\end{abstract}



\section{Introduction}

The putative dust torus is an important ingredient of the
unification model for active galactic nuclei (AGNs) \citep{a93}. A
type 1 AGN spectrum is yielded if the broad-line region (BLR) is
directly viewed through the hole of the face-on dust torus, while
the BLR is obscured by the torus seen edge-on that leads to a type
2 AGN spectrum. This model can successfully unify Seyfert 1 and
Seyfert 2 galaxies; radio quasars and radio galaxies, etc (e.g.,
Antonucci \& Miller, 1985; Miller \& Goodrich 1990; Tran 1995;
Barthel 1989), which is supported by a variety of observational
features of AGNs (e.g., Lawrence 1991; Hill, Goodrich, \& Depoy,
1996).

The dust torus is irradiated by the central engine of the AGN, and
the dust cannot survive inside a critical radius at which the
temperature is so high that the dust begins to sublimate. The
temperature of the dust clouds at the inner edge of the torus is
close to the sublimation temperature. The irradiated dust produces
thermal emission mostly at the NIR wavelength. The decomposition
of spectral energy distributions of AGNs suggested that the NIR
continuum emission ($\ga2{\rm\mu{m}}$) from Seyfert galaxies is
dominated by thermal radiation from the hot dust surrounding the
central engine (e.g., Kobayashi et al. 1993). The analysis on the
infrared emission data of 64 Palomar-Green(PG) quasars favors the
dust being heated by the central engines of AGNs \citep{h03}. The
inner radius of the dust torus is assumed to be roughly at the
dust evaporation radius $R(T_{\rm evap})\simeq 0.06(L_{\rm
bol}/10^{38}{\rm W})^{1/2}$ pc \citep{nl93,hgd96}. The intense
monitoring observations were carried out on the Seyfert 1 galaxy
NGC4151, and a lag time of $48^{+2}_{-3}$ days between the $V$ and
$K$ light curves was measured. The inner radius of the dust torus
in NGC4151 is $\sim 0.04$ pc corresponding to this measured time
lag. For this source, the predicted evaporation radius $R(T_{\rm
evap})\simeq 0.015$ pc (Netzer \& Laor, 1996; $L_{\rm bol}\simeq
6.5\times 10^{36}$ W, given by Kaspi et al., 2000), which is
roughly consistent with the inner radius of the dust torus $\sim
0.04$ pc measured by thermal dust reverberation method
\citep{m04}. If the radio quasars and galaxies are intrinsically
same and can be unified by their different orientations,  the
relative thickness of the torus $H/R\sim 2-3$ is estimated from
the fraction of quasars in the 3CR sample \citep{hgd96}.

If the relative thickness $H/R$ of the torus is around unity, the
random velocities should be $\ga$100km~s$^{-1}$ for the matter in
the tori at sub-parsec scales, because the dust torus is required
to be in dynamical equilibrium.  If this motion is thermal, the
corresponding temperature ($\sim$10$^6$~K) is too high for dust to
survive. It is therefore suggested that the cold dust is in the
clouds moving at the random velocities $\ga$100km~s$^{-1}$
\citep{kb88}. The origin of such dust tori in AGNs is still a
mystery, though \citet{zb01,zb02} suggested that the merger of
binary black holes may lead the distribution of the surrounding
matters to be a torus-like structure. In this paper, we perform
statistic analysis on the properties of the dust tori in PG
quasars. The cosmological parameters $\Omega_{\rm M}=0.3$,
$\Omega_{\Lambda}=0.7$, and $H_0=70~ {\rm km~s^{-1}~Mpc^{-1}}$
have been adopted in this work.

\section{Black hole mass}

\citet{k00} derived an empirical relation between the BLR size and
optical continuum luminosity for a sample of Seyfert 1 galaxies
and quasars using the cosmological parameters $H_0=75~ {\rm
km~s^{-1}~Mpc^{-1}}$ and $q_0=0.5$, in which the sizes of BLRs are
measured with the reverberation mapping method \citep{p93}. The
relation \be R_{\rm BLR}=(26.4\pm 3.0)\left[ {\frac {\lambda
L_{\lambda}(5100 {\AA})}{10^{44}{\rm ergs~
s^{-1}}}}\right]^{0.610\pm0.10} {\rm lt-days} \label{rblr} \ee is
derived by \citet{mj02} for the same sample, but the cosmological
parameters used are same as this work: $\Omega_{\rm M}=0.3$,
$\Omega_{\Lambda}=0.7$, and $H_0=70~ {\rm km~s^{-1}~Mpc^{-1}}$.

The central black hole masses $M_{\rm bh}$ can be estimated from
the velocities $v_{\rm BLR}$ of the clouds in the BLRs \be M_{\rm
bh}={\frac {v_{\rm BLR}^2R_{\rm BLR}}{G}}, \label{mbh}\ee where
the motions of the clouds are assumed to be virilized and
isotropic. The velocities of the clouds in BLRs $v_{\rm BLR}$ are
derived from the width of the broad emission lines.  For most
quasars, the BLR sizes have not been measured by the reverberation
mapping method, and the empirical relation (\ref{rblr}) is instead
used to estimate the BLR sizes. The central black hole masses of
quasars have been estimated from their broad-line widths and
optical continuum luminosity (e.g., Laor, 2000; McLure \& Dunlop
2001; Cao \& Jiang 2002). For normal bright quasars, the
bolometric luminosity can be estimated from their optical
luminosity $L_{\lambda,\rm opt}$ at 5100 $\AA$ by \citep{k00} \be
L_{\rm bol}\simeq 9\lambda L_{\rm \lambda,opt}. \label{lbolopt}\ee
Assuming a constant accretion efficiency for all quasars, we have
the conventional defined dimensionless accretion rate \be
\dot{m}={\frac {L_{\rm bol}}{L_{\rm Edd}}}. \label{mdot}\ee

\section{Torus geometry}

The dust clouds at the inner edge of the torus irradiated by the
AGN central engine produce thermal emission mostly in the NIR
waveband (e.g., Kobayashi et al., 1993), so the NIR luminosity of
the dust torus is estimated by \be L_{\rm NIR}\simeq {\frac
{\Delta\Omega_{\rm torus}(R_{\rm in})}{4\pi}}f_{\rm in}L_{\rm
bol}, \label{domega}\ee where $L_{\rm bol}$ is the bolometric
luminosity of the AGN, $\Delta\Omega_{\rm torus}$ is the solid
angle subtended by the dust torus as seen from the central source
at the inner radius $R_{\rm in}$ of the torus, and the factor
$f_{\rm in}$ describes the covering factor of the clouds at the
inner radius of the torus and the uncertainties. The uncertainties
arising from the NIR emission contributed by the central continuum
emission of the AGN and starbursts in the AGN host galaxy may not
be large, because this emission can be neglected compared with the
NIR emission from the tori in most PG quasars \citep{h03}. {If the
contribution in the NIR waveband from stars in the host galaxy can
be neglected compared with that from the torus, the factor $f_{\rm
in}$ describes the covering factor.} Thus, we can infer the dust
torus geometry from the NIR and bolometric luminosities of AGNs,
\be \Delta\Omega_{\rm torus}(R_{\rm in})=4\pi f_{\rm
in}^{-1}{\frac {L_{\rm NIR}}{L_{\rm bol}}}, \label{domega1}\ee if
the covering factor of the dust clouds at the inner edge of the
torus is known. If we assume all the IR emission of quasars is
from the dust torus irradiated by the AGNs, we can estimate the
solid angle subtended by the whole dust torus by \be
\Delta\Omega_{\rm torus}^{\rm max}=4\pi {\frac {L_{\rm IR}}{L_{\rm
bol}}}, \label{domega2}\ee where the covering factor $f=1$ is
adopted, because the putative torus is required to be able to
obscure the nuclear emission while it is seen edge-on \citep{a93}.
If the torus is subtended at a constant solid angle with radius
(i.e., cone-like structure), the covering factor $f_{\rm in}$ at
the inner radius $R_{\rm in}$ can be estimated by \be f_{\rm
in}\simeq {\frac {L_{\rm NIR}}{L_{\rm IR}}}.\label{fin}\ee The
solid angle estimated from Eq. (\ref{domega2}) is an upper limit
for some sources of which the IR emission from the host galaxies
is important compared with that from the dust tori.

The solid angle subtended by the torus $\Delta\Omega_{\rm torus}$
is  \be \Delta\Omega_{\rm torus}={\frac {4\pi
H/R}{(4+H^2/R^2)^{1/2}}}, \label{domegah} \ee where $H$ is the
thickness of the torus. For the case of $H/R\la 1$, Equation
(\ref{domegah}) can be approximated as \be \Delta\Omega_{\rm
torus}\simeq {\frac {2\pi H}{R}}. \label{domegah1} \ee Combining
the relations (\ref{domega1}) and (\ref{domegah}), we can roughly
estimate the torus thickness at its inner edge by \be H(R_{\rm
in})={\frac {2f_{\rm in}^{-1}R_{\rm in}(L_{\rm NIR}/L_{\rm bol})}
{[1-f_{\rm in}^{-2}(L_{\rm NIR}/L_{\rm bol})^2]^{1/2}}} ~~~{\rm
pc}, \label{finh} \ee where the inner radius $R_{\rm in}$ of the
torus is given by $R_{\rm in}\simeq 0.06(L_{\rm bol}/10^{38}{\rm
W})^{1/2}$ pc \citep{nl93,hgd96}.

\section{Sample}

\citet{h03} provided a sample of 64 PG quasars with infrared
spectral energy distributions ($3-150{\rm {\mu}m}$) observed by
the  ISO. The PG quasars have high infrared detection rate of more
than 80 per cent. As discussed in Sect. 2, the central black hole
masses in quasars can be estimated from their broad-line widths of
H$\beta$ and optical continuum luminosities $\lambda
L_{\lambda}(5100 {\AA})$. The full widths at half maximum (FWHM)
of broad-line H$\beta$ for 51 sources in this sample are available
in \citet{bg92}. For the remainder, they are at relatively high
redshifts ($z\ga 1$). We search the literature and find
FWHM(H$\beta$)=5100~km~s$^{-1}$ for 0044$+$030 \citep{b96}; and
FWHM(H$\beta$)=9601 ~km~s$^{-1}$, 4613 ~km~s$^{-1}$ for 1634$+$706
and 1718$+$481, respectively \citep{s03}.  It leads to a sample of
54 quasars with estimated black hole masses, which includes 41
radio-quiet quasars and 13 radio-loud quasars (4 flat-spectrum and
9 steep-spectrum radio quasars). There are ten sources with
FWHM(H$\beta$)$<2000$~km~s$^{-1}$ in this sample (hereafter we
refer to those quasars with FWHM(H$\beta$)$\ge2000$~km~s$^{-1}$ as
broad-line(BL) quasars).

\section{Results}

We estimate the central black hole masses of these PG quasars
using their broad-line widths of H$\beta$ and optical continuum
luminosities as described in Sect. 2. The NIR luminosities $L_{\rm
NIR}$ of the PG quasars in the waveband of $3-10~{\mu{\rm m}}$
given by \citet{h03} are used in this work to explore the
properties of the tori at their inner radii, because the emission
from the inner region of the dust torus irradiated by the AGN is
dominant in the NIR waveband. This is supported by the thermal
dust reverberation measurements on NGC4151 \citep{m04}. In Fig.
\ref{fig1}, we plot the relation between the ratios $L_{\rm
bol}/L_{\rm Edd}$ and $L_{\rm NIR}/L_{\rm bol}$. No correlation is
found between these two ratios.

\citet{h03} suggested a scheme of quasar evolution for their dust
distribution surrounding the AGNs. The sources are divided into
different classes according to their evolutionary sequence. The
classes 0 and 1 represent cool and warm ultra-luminous infrared
galaxies respectively. Class 2 sources are young quasars, and the
starbursts are still important in IR wavebands compared with the
torus emission. There are two class 2 quasars in this sample, and
we leave out these two infrared luminous quasars $0157+001$ and
$1351+640$ in all our statistic analysis, because the contribution
in the infrared waveband by the dust heated by the stars in these
two quasars may not be neglected, though these two sources would
only affect little on the statistic results.
The relation between the black hole mass $M_{\rm bh}$ and the
ratio $L_{\rm NIR}/L_{\rm bol}$ is plotted in Fig. \ref{fig2}. The
generalized Kendall's $\tau$ test (Astronomical Survival Analysis,
ASURV, Feigelson \& Nelson 1985) shows that the correlation is
significant at the 99.93 per cent level for the whole sample of 52
sources (the significant level is almost same at the 99.94 per
cent for the sample of 42 BL quasars). The linear regression by
parametric expectation-maximization (EM) Algorithm (ASURV) gives
\be \log_{10} L_{\rm NIR}/L_{\rm bol}
=-0.097(\pm0.034)\log_{10}M_{\rm bh}/{\rm
M}_{\odot}-0.125(\pm0.279), \ee for the whole sample (the solid
line in Fig. \ref{fig2}), and \be \log_{10} L_{\rm NIR}/L_{\rm
bol} =-0.149(\pm0.038)\log_{10}M_{\rm bh}/{\rm
M}_{\odot}+0.325(\pm0.322), \ee for the broad-line quasars (the
dotted line in Fig. \ref{fig2}), respectively. Compared with the
linear regression for the whole sample, the slope becomes slightly
steeper for the BL quasars, while the correlation levels are
almost same. For those 13 radio-loud quasars in this sample, no
statistic behavior systematically different from the radio-quiet
quasars is found.

The BLR size $R_{\rm BLR}$ is derived from the optical continuum
luminosity at 5100 $\AA$, so the estimated black hole mass $M_{\rm
bh}\propto v_{\rm BLR}^2L^{0.61}_{\lambda}(5100\AA)$ is given by
substituting Eq. (\ref{rblr}) into Eq. (\ref{mbh}). The bolometric
luminosity $L_{\rm bol}$ is estimated from the optical continuum
luminosity by using the empirical relation (\ref{lbolopt}).  It is
therefore doubtful that the correlation between the black hole
mass $M_{\rm bh}$ and the ratio $L_{\rm NIR}/L_{\rm bol}$ may be
caused by the common dependence of optical continuum luminosity
$L_\lambda$. We plot the relation between the bolometric
luminosity $L_{\rm bol}$ and the ratio $L_{\rm NIR}/L_{\rm bol}$
in Fig. \ref{fig3}. It is found that the bolometric luminosity
$L_{\rm bol}$ is correlated with the ratio $L_{\rm NIR}/L_{\rm
bol}$ for the whole sample.  In order to test the correlation
between the black hole mass  $M_{\rm bh}$ and the ratio $L_{\rm
NIR}/L_{\rm bol}$ is affected to what extent by the common
dependence of the optical continuum luminosity, we perform
statistic analysis on the sources with $10^{45}~{\rm
erg~s^{-1}}\le L_{\rm bol}\le 10^{46.5}~{\rm erg~s^{-1}}$ (35
sources situated between the two dotted lines in Fig. \ref{fig3}).
We find that the ratio $L_{\rm NIR}/L_{\rm bol}$ is still
significantly correlated with the black hole mass $M_{\rm bh}$ at
the 97.25 per cent level for this sub-sample of 35 quasars, {
while no correlation is found between $L_{\rm bol}$ and $L_{\rm
NIR}/L_{\rm bol}$ (at the 75.01 per cent level).} This means that
the ratio $L_{\rm NIR}/L_{\rm bol}$ is intrinsically correlated
with the black hole mass $M_{\rm bh}$.

We also perform statistic analysis on the relations: $M_{\rm
bh}-L_{\rm MIR}/L_{\rm bol}$, $M_{\rm bh}-L_{\rm FIR}/L_{\rm
bol}$, and $M_{\rm bh}-L_{\rm IR}/L_{\rm bol}$ (see Figs.
\ref{fig4}-\ref{fig6}). The mid-infrared (MIR,
$10-40~{\rm\mu{m}}$), far-infrared (FIR, $40-150~{\rm\mu{m}}$),
and total infrared (IR, $3-150~{\rm\mu{m}}$) luminosities of these
quasars are taken from \citet{h03}. It is found that they are
significantly correlated at the similar levels as the NIR case,
while the slope becomes steeper for the FIR case. There are three
sources in our sample have $L_{\rm IR}/L_{\rm bol}\ga 1$ (see Fig.
\ref{fig6}). The detailed statistic results are summarized in
Table 1.

The dust clouds at the inner radius $R_{\rm in}$ of the torus
irradiated by the AGN emit mostly in the NIR waveband. The
covering factor $f_{\rm in}$ at $R_{\rm in}$ can be estimated by
using the relation (\ref{fin}). Subtracting the three sources with
$L_{\rm IR}/L_{\rm bol}\ga 1$ and the infrared luminous source
$1351+640$, we find that the ratios $L_{\rm NIR}/L_{\rm IR}$  for
the sources in our present sample are in the range $0.20-0.56$
with an average of 0.39. We use Eq. (\ref{finh}) to estimate the
torus thickness at its inner radius $R_{\rm in}$ adopting a
covering factor $f_{\rm in}=0.39$.
The relations of $M_{\rm bh}-H(R_{\rm in})$ and $L_{\rm
bol}-H(R_{\rm in})$ are plotted in Figs. \ref{fig7} and \ref{fig8}
respectively. It is found that the torus thickness $H$ at its
inner radius $R_{\rm in}$ increases with the black hole mass
$M_{\rm bh}$ and the bolometric luminosity $L_{\rm bol}$.

The solid angle $\Delta\Omega_{\rm torus}$ subtended by the whole
torus can be estimated by using Eq. (\ref{domega2}) from the ratio
$L_{\rm IR}/L_{\rm bol}$. We can then calculate the relative torus
thickness $H/R$ using Eq. (\ref{domegah}). In Fig. \ref{fig9}, we
plot the relation between the black hole mass $M_{\rm bh}$ and the
relative torus thickness $H/R$. The three sources with $L_{\rm
IR}/L_{\rm bol}\ga 1$ do not appear in this figure. The statistic
results are listed in Table 1.

\figurenum{1}
\centerline{\includegraphics[angle=0,width=10.0cm]{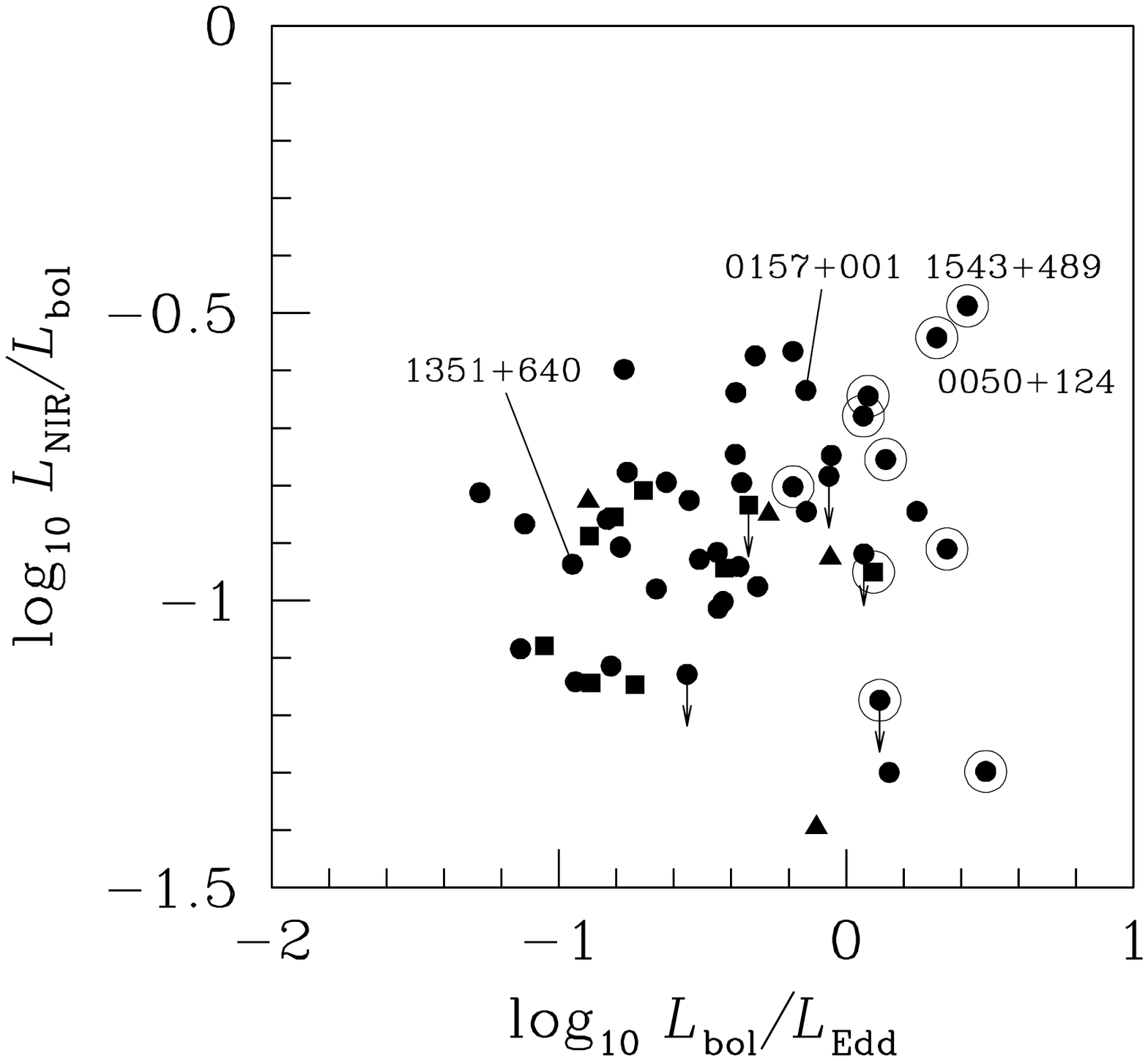}}
\figcaption{\footnotesize The relation between the ratios $L_{\rm
bol}/L_{\rm Edd}$ and $L_{\rm NIR}/L_{\rm bol}$. The circles
represent radio-quiet quasars, while the squares and triangles
represent steep-spectrum and flat-spectrum radio quasars,
respectively. The sources with large circles have
FWHM(H$\beta$)$<$2000~km~s$^{-1}$. The relation between the ratios
$L_{\rm bol}/L_{\rm Edd}$ and $L_{\rm NIR}/L_{\rm bol}$.
\label{fig1}} \centerline{}

\figurenum{2}
\centerline{\includegraphics[angle=0,width=10.0cm]{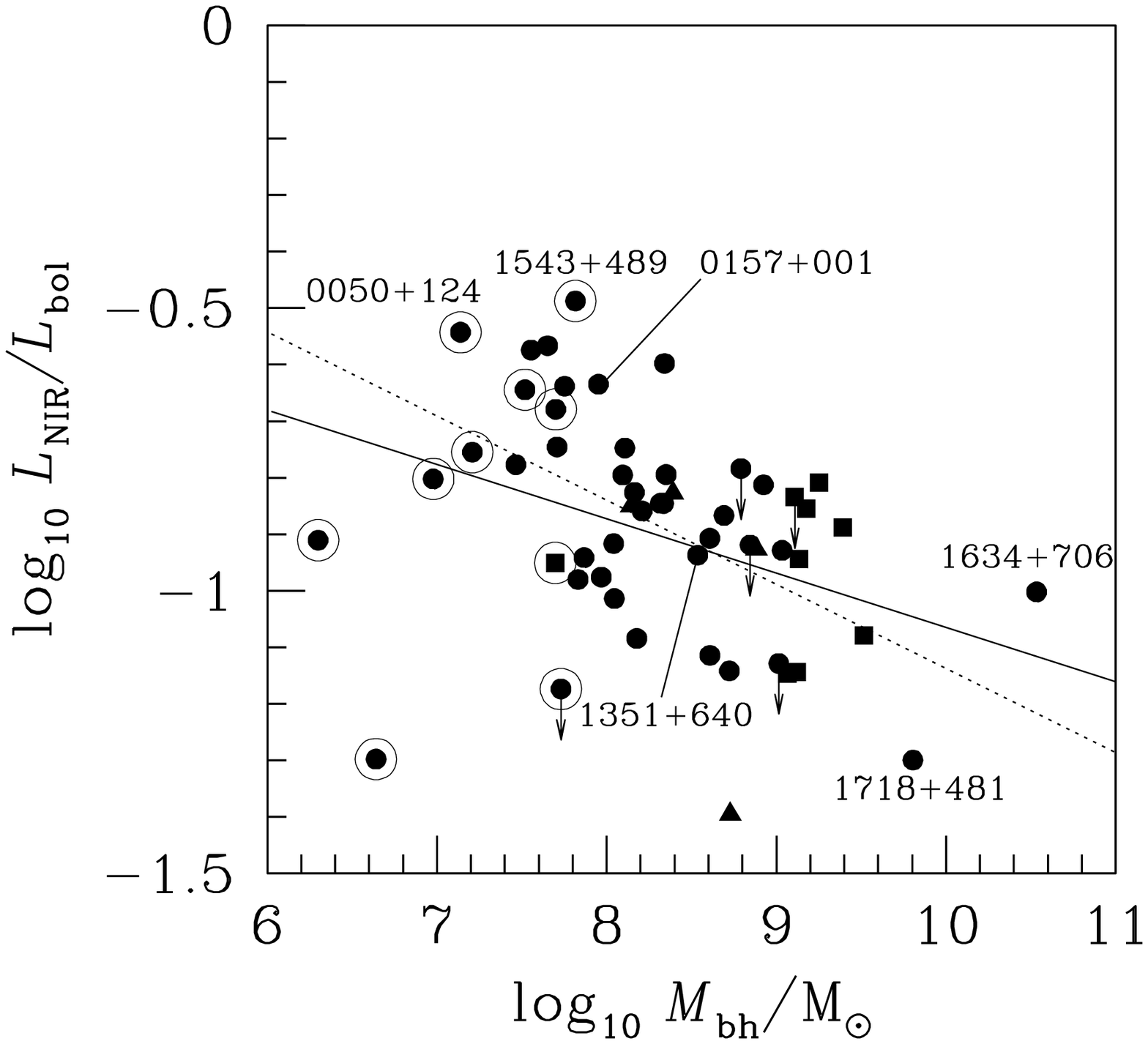}}
\figcaption{\footnotesize The relation between the black hole mass
$M_{\rm bh}$  and the ratio $L_{\rm NIR}/L_{\rm bol}$. The solid
line represents the linear regression for the whole sample but
excluding two infrared luminous sources $0157+001$ and $1351+640$.
The dotted line represents the linear regression for the sources
with FWHM(H$\beta$)$\ge 2000$~km~s$^{-1}$ excluding two sources
$0157+001$ and $1351+640$.  \label{fig2}} \centerline{}

\figurenum{3}
\centerline{\includegraphics[angle=0,width=10.0cm]{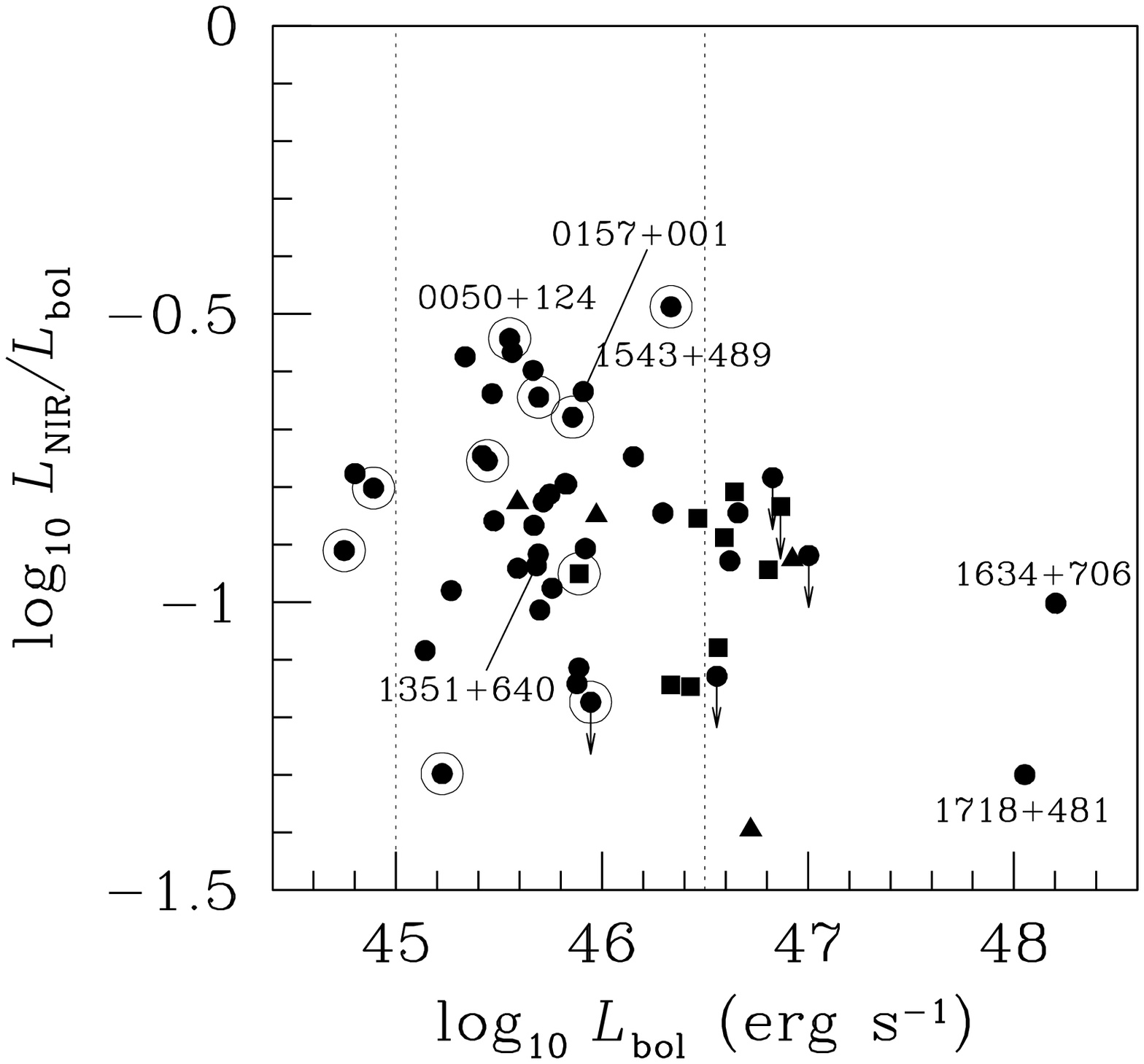}}
\figcaption{\footnotesize The relation between the ratios $L_{\rm
bol}$ and $L_{\rm NIR}/L_{\rm bol}$. The dotted lines represent
$L_{\rm bol}=10^{45}$~erg~s$^{-1}$ and $10^{46.5}$~erg~s$^{-1}$.
\label{fig3}} \centerline{}

\figurenum{4}
\centerline{\includegraphics[angle=0,width=10.0cm]{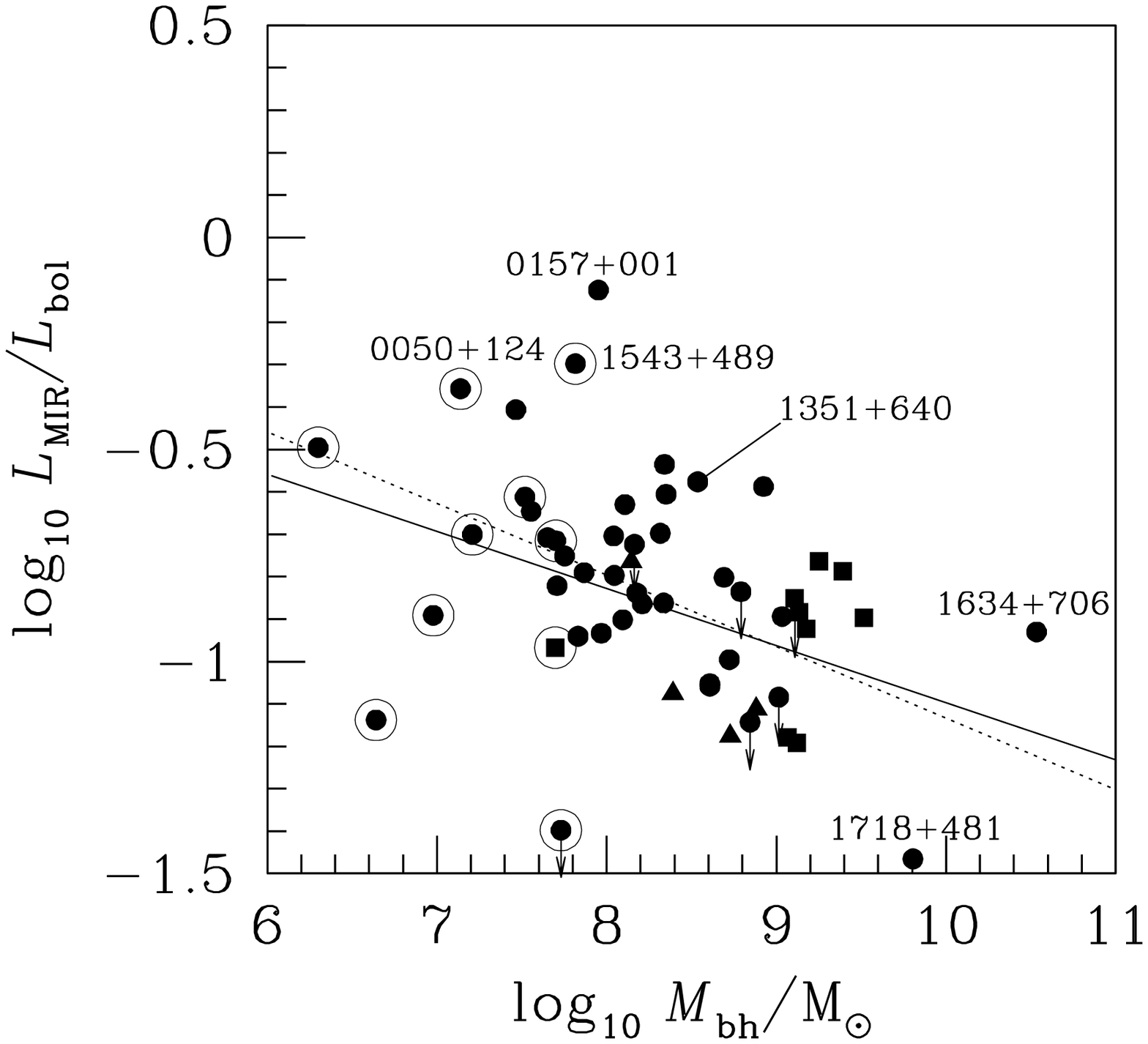}}
\figcaption{\footnotesize Same as Fig. \ref{fig2}, but for the
ratio $L_{\rm MIR}/L_{\rm bol}$.    \label{fig4}} \centerline{}

\figurenum{5}
\centerline{\includegraphics[angle=0,width=10.0cm]{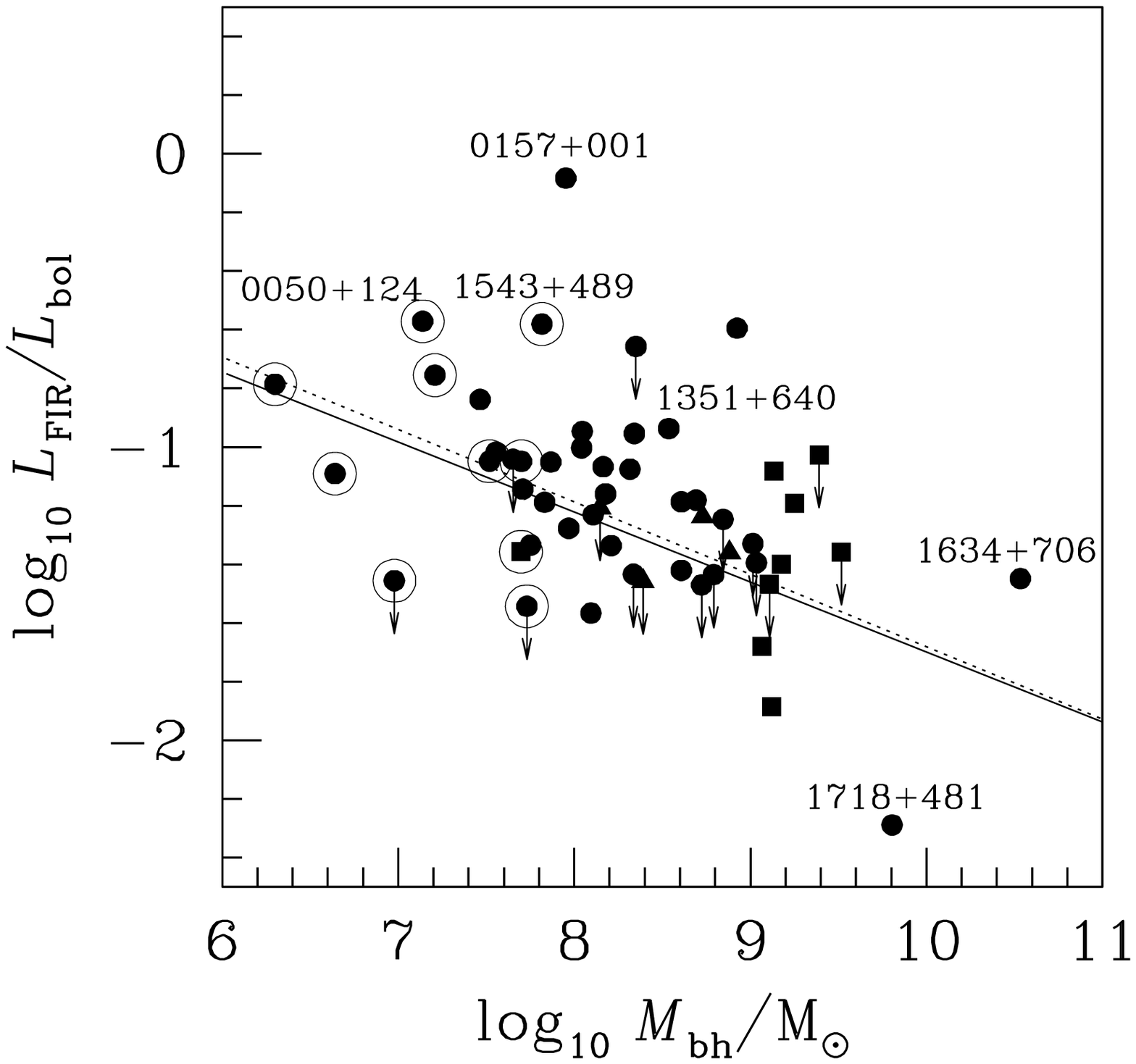}}
\figcaption{\footnotesize Same as Fig. \ref{fig2}, but for the
ratio $L_{\rm FIR}/L_{\rm bol}$.    \label{fig5}} \centerline{}

\figurenum{6}
\centerline{\includegraphics[angle=0,width=10.0cm]{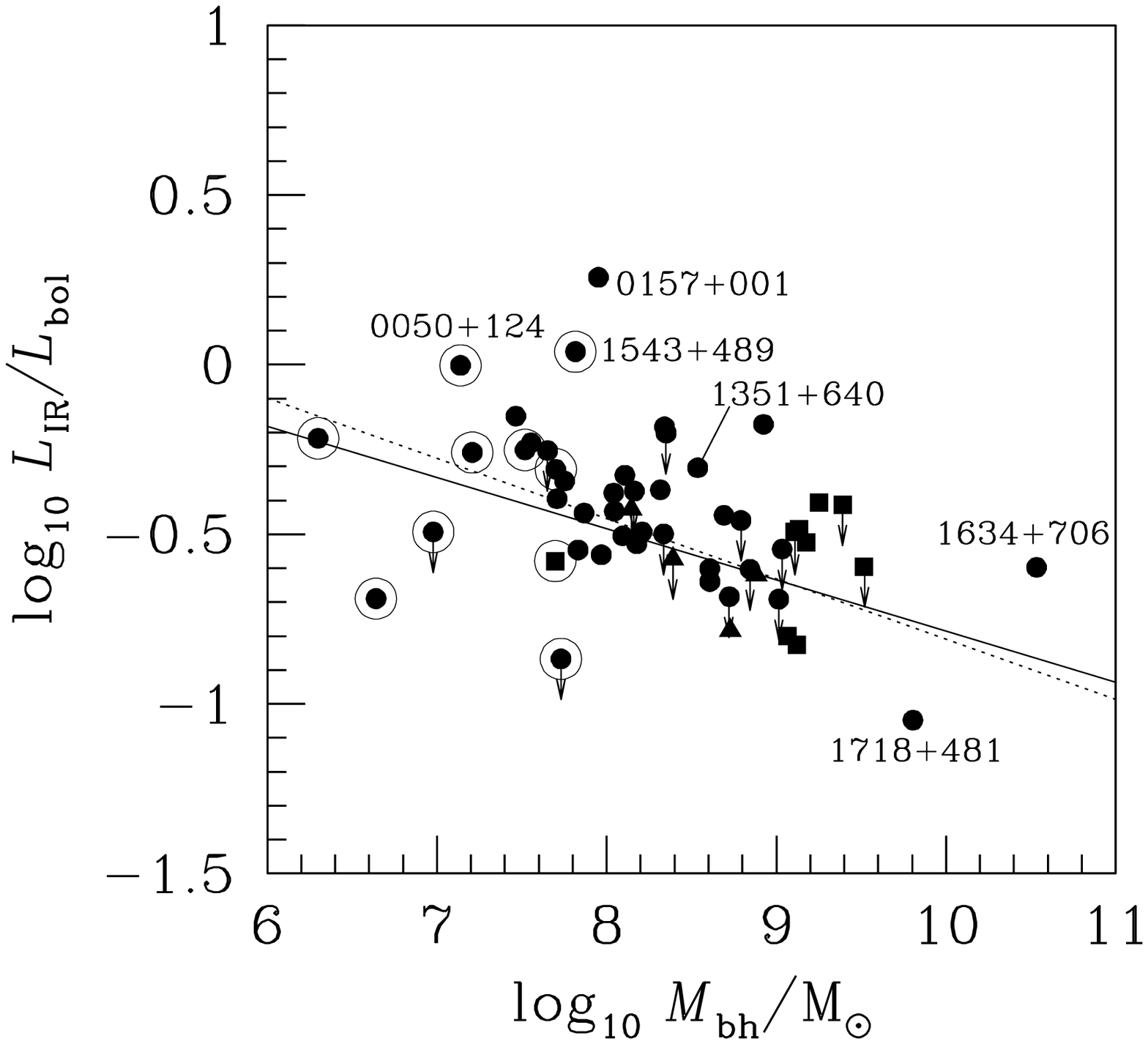}}
\figcaption{\footnotesize Same as Fig. \ref{fig2}, but for the
ratio $L_{\rm IR}(3-150\mu{\rm m})/L_{\rm bol}$.   \label{fig6}}
\centerline{}

\figurenum{7}
\centerline{\includegraphics[angle=0,width=10.0cm]{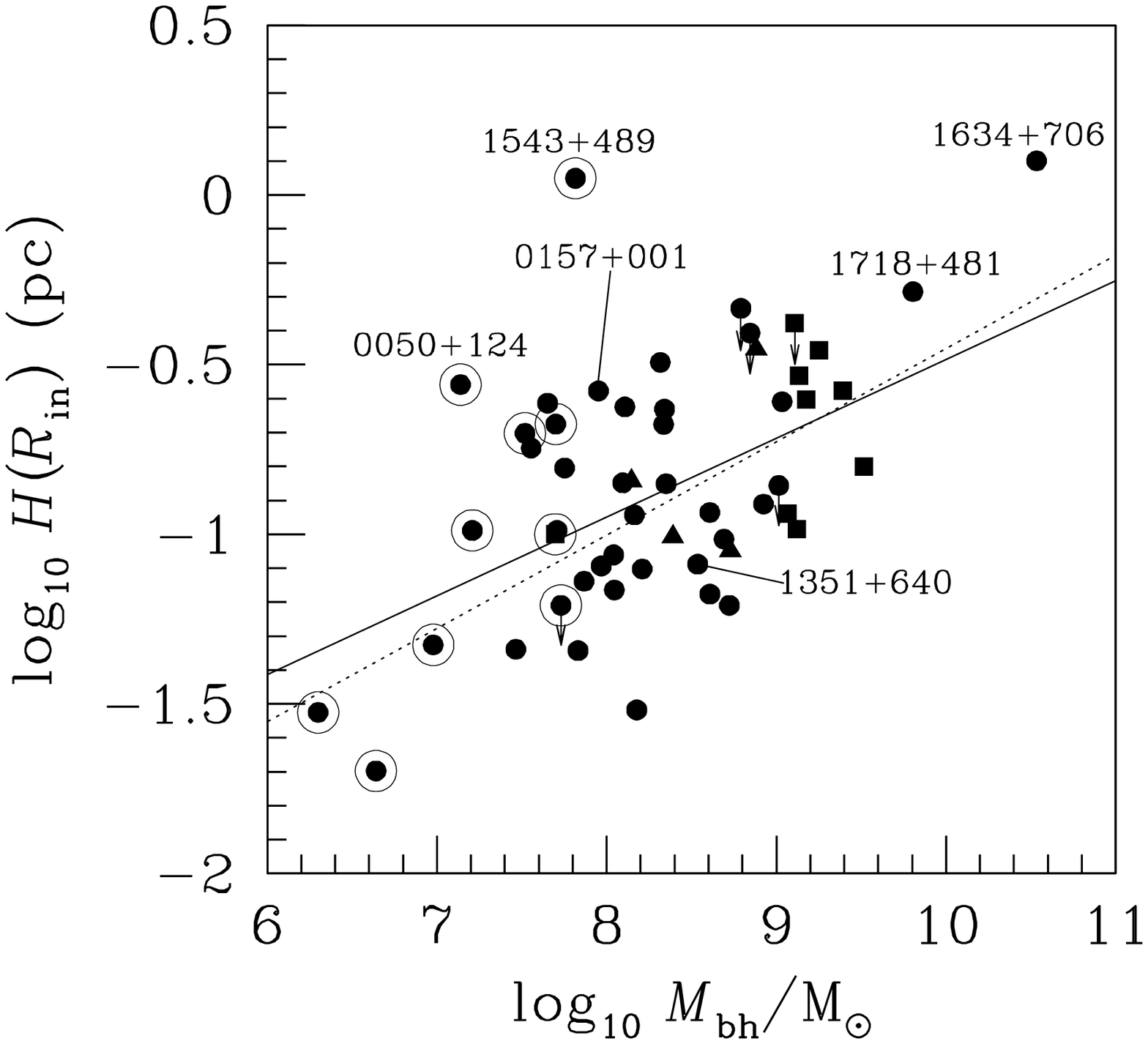}}
\figcaption{\footnotesize The relation between the black hole mass
$M_{\rm bh}$ and the torus thickness $H(R_{\rm in})$ at its inner
radius $R_{\rm in}$.
The solid line represents the linear regression for the whole
sample but excluding the infrared luminous sources $0157+001$ and
$1351+640$. The dotted line represents the linear regression for
the sources with FWHM(H$\beta$)$\ge 2000$~km~s$^{-1}$ excluding
the sources $0157+001$ and $1351+640$.  \label{fig7}}
\centerline{}

\figurenum{8}
\centerline{\includegraphics[angle=0,width=10.0cm]{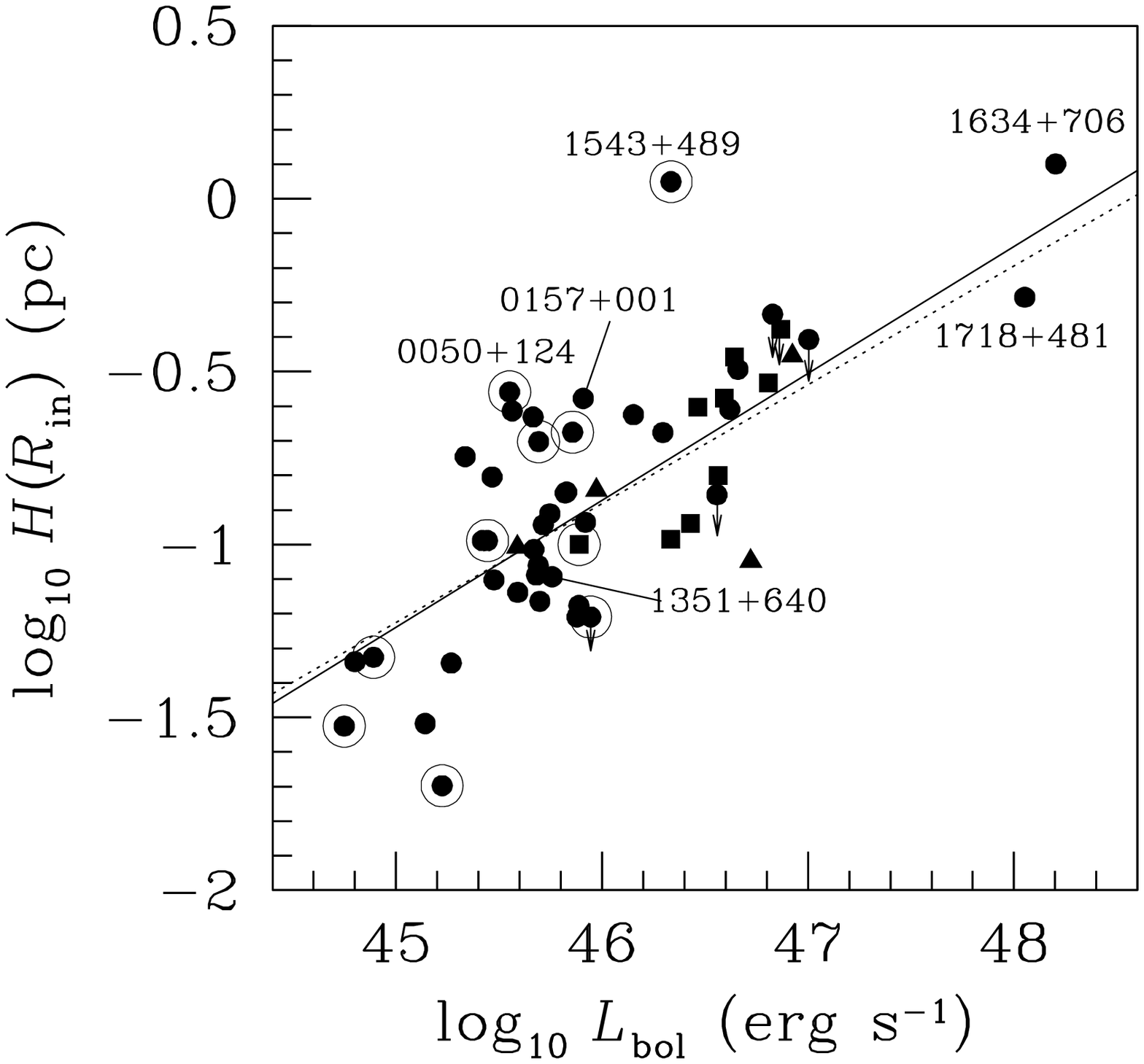}}
\figcaption{\footnotesize The relation between the bolometric
luminosity $L_{\rm bol}$ and the torus thickness $H(R_{\rm in})$
at its inner radius $R_{\rm in}$. The solid line represents the
linear regression for the whole sample but excluding the infrared
luminous sources $0157+001$ and $1351+640$. The dotted line
represents the linear regression for the sources with
FWHM(H$\beta$)$\ge 2000$~km~s$^{-1}$ excluding the sources
$0157+001$ and $1351+640$.\label{fig8}} \centerline{}

\figurenum{9}
\centerline{\includegraphics[angle=0,width=10.0cm]{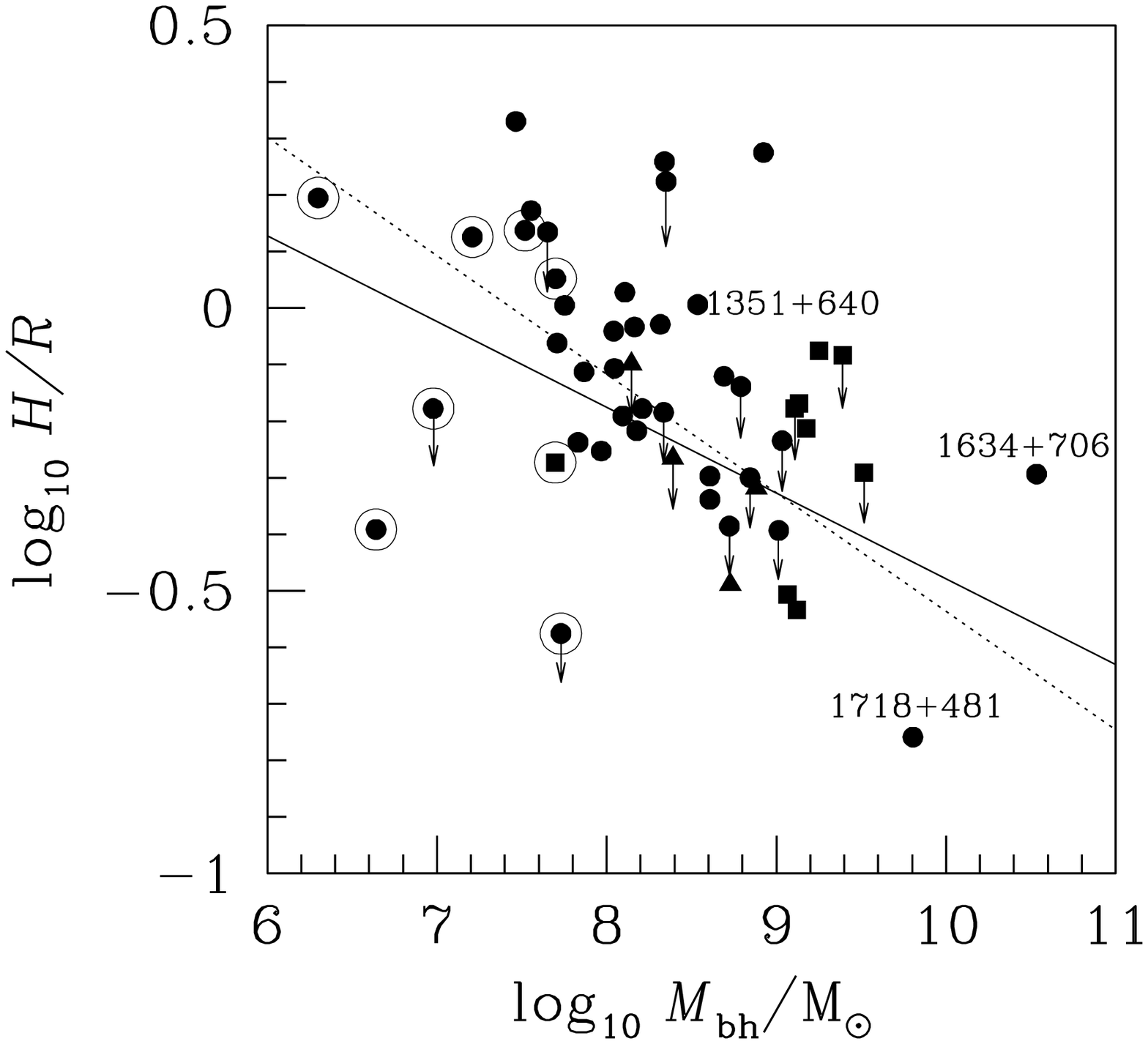}}
\figcaption{\footnotesize The relation between black hole mass
$M_{\rm bh}$ and the relative thickness of the torus $H/R$.
\label{fig9}} \centerline{}

\section{Discussion}

The NIR luminosities $L_{\rm NIR}$ of the PG quasars in the
waveband of $3-10~{\mu{\rm m}}$ given by \citet{h03} are used in
this work. The NIR emission in the waveband $\ga 2~\mu{\rm m}$ is
dominated by the thermal emission from the dust torus, which is
supported by the decomposition of the SEDs of the Seyfert galaxies
and the thermal dust reverberation measurements on NGC4151
\citep{k93,m04}. The NIR emission $\sim 2{\rm\mu{m}}$ is not
included in the NIR luminosities $L_{\rm NIR}$ provided by
\citet{h03}, so the solid angle $\Omega_{\rm torus}$ subtended by
the torus may be slightly larger than that estimated from the
ratio $L_{\rm NIR}/L_{\rm bol}$ by using the relation
(\ref{domega1}), No correlation is found between the ratios
$L_{\rm bol}/L_{\rm Edd}$ and $L_{\rm NIR}/L_{\rm bol}$, which
implies that the solid angle $\Delta\Omega_{\rm torus}$ subtended
by the dust torus does not evolve with the accretion rate
$\dot{m}$.
In the clumpy dust cloud model \citep{kb88}, the dimensionless
torus thickness is estimated by \be {\frac {H}{R}}\sim {\frac
{2V_{\rm c}}{V_{\rm K}}}, \ee where $V_{\rm c}$ is the typical
random velocity of the dust clouds and $V_{\rm K}$ is the
Keplerian velocity at radius $R$. This means the ratio of the
typical dust cloud random velocity to the Keplerian velocity
$V_{\rm c}/V_{\rm K}$ is independent of the dimensionless
accretion rate $\dot{m}$. However, it should be cautious that the
PG quasars are optically selected and they are very bright (most
sources have $L_{\rm bol}/L_{\rm Edd}\ga 0.1$, see Fig.
\ref{fig1}). We cannot rule out the presence of a correlation
between the ratios $L_{\rm bol}/L_{\rm Edd}$ and $L_{\rm
NIR}/L_{\rm bol}$, if more fainter quasars with lower values of
$L_{\rm bol}/L_{\rm Edd}$ are included in the statistic analysis.

The intrinsic correlation between black hole mass $M_{\rm bh}$ and
the ratio $L_{\rm NIR}/L_{\rm bol}$ indicates that the dust torus
geometry is related with the central black hole mass. The larger
black hole leads to a dust torus with smaller solid angle
subtended by it seen from the central source. The black hole grows
slowly during its bright quasar phase (usually less than a factor
of 2, see Kauffmann \& Haehnelt, 2000), because the typical
timescale for the bright quasar phase is $\sim 10^{7-8}$ years
(e.g., Yu \& Tremaine, 2002). This means that the black hole mass
estimated from the broad-line width and the optical continuum
luminosity can roughly reflect the hole mass while the bright
quasar was formed at an early time. If the dust torus was formed
at nearly the same time as the bright quasar,  it may imply that
the formation of the dust torus is regulated by the central black
hole, though the detailed mechanism is still unclear.
\citet{zb01,zb02} suggested a torus formation scenario that the
merger of binary black holes may lead the distribution of the
surrounding matter to be a torus-like structure. In this scenario,
the black holes play important roles in the formation of the
torus, which is in general consistent with our statistic results.

Similar correlations between the black hole mass $M_{\rm bh}$ and
the ratio $L_{\rm IR}/L_{\rm bol}$ are also found for the MIR and
FIR cases (see Fig. \ref{fig4} and \ref{fig5}). This implies that
the emission in the MIR and FIR wavebands is dominated by the
thermal dust tori heated by the central AGNs, excepting the
infrared luminous quasars $0157+001$. This source has very high
ratios of the infrared to bolometric luminosities, especially in
the MIR and FIR wavebands, which implies that the starbursts in
this source contribute much in the MIR and FIR wavebands than the
NIR waveband. It is still unclear why the slope of the correlation
between $M_{\rm bh}$ and $L_{\rm FIR}/L_{\rm bol}$ is steeper than
that for the NIR and MIR cases (see Table 1 for comparison). One
likely explanation is that the torus properties (e.g., number of
the clouds and their distribution, and the cloud optical depth,
etc.) or/and the AGN spectral shape vary systematically with the
black hole mass $M_{\rm bh}$, because the spectrum emitted from
the dust torus is determined by the torus properties, and the SED
of the AGN (e.g., Nenkova, Ivezic, \& Elitzur, 2002). The detailed
model calculations, for example, with the code DUSTY
\citep{nie02}, are necessary for solving this issue, which are
beyond the scope of this work. An alternative explanation is that
the starbursts contribution in the FIR waveband is more important
than the NIR and MIR wavebands, and the thermal emission from the
dust torus irradiated by the AGN contributes less in the FIR
waveband. In this case,  the starbursts contribution is supposed
to change the slope of the correlation between $M_{\rm bh}$ and
$L_{\rm FIR}/L_{\rm bol}$, while it should still not be dominant
to avoid destroying this correlation. This can be explained by
recent IR observations showing that the nuclear starbursts  are in
the dusty tori around the nuclei of some  Seyfert galaxies
\citep{iw04}.

For most quasars in our sample, the IR emission is dominantly from
the tori irradiated by the AGNs. In this case, the IR luminosity
should be less than the bolometric luminosity. In Fig. \ref{fig6},
we find three sources with $L_{\rm IR}(3-150\mu{\rm m})/L_{\rm
bol}\ga 1$. The IR emission from these four sources (including one
class 2 quasars: 0157$+$001; two narrow-line quasars: 0050$+$124
and 1543$+$489) cannot be solely produced by the dust torus heated
by the AGNs. The contribution of stars in the host galaxies should
be important in these four quasars. The ratio $L_{\rm IR}/L_{\rm
bol}\ga 1$ can therefore be used as a criterion for starbursts
dominant quasars. The detailed geometry of the dust torus is still
unclear. The opening angle of the torus may vary with radius.
However, the ratio $L_{\rm IR}/L_{\rm bol}$ can always be used to
estimate the solid angle subtended by the whole torus. For those
four quasars with $L_{\rm IR}/L_{\rm bol}>1$, the IR emission is
dominated by the starbursts and we are not able to estimate the
solid angle subtended by the torus from the ratio $L_{\rm
IR}/L_{\rm bol}$. We find that the relative torus thickness $H/R$
estimated from the ratio $L_{\rm IR}/L_{\rm bol}$ is in the range
of $0.2-2.4$ with an average of $\sim 0.9$ (see Fig. \ref{fig9}).
This is different from $H/R\sim 2-3$ estimated from the fraction
of quasars in the 3CR sample \citep{hgd96}.

We may estimate the solid angle $\Delta\Omega_{\rm torus}$
subtended by the torus at its inner radius from the ratio of the
NIR to bolometric luminosities, because the NIR emission is
dominated by the emission from the dust clouds at its inner radius
irradiated by the AGN. The emission in the NIR waveband is less
affected by the starbursts compared with the MIR and FIR
wavebands. If the opening angle is constant for the torus at
different radius, the covering factor of the clouds at the inner
radius can be estimated from the ratio $L_{\rm NIR}/L_{\rm IR}$
(Eq. \ref{fin}). It implies that only a fraction of the nuclear
radiation is absorbed by the dust clouds at the inner radius of
the torus. The remainder may be absorbed by the dust clouds at
larger radii, and they have lower temperatures than that at the
inner radius. Most of their emission may be in the MIR or FIR
wavebands depending on their distance from the nucleus. This is
consistent with the fact that similar correlations are found
between $M_{\rm bh}$ and $L_{\rm IR}/L_{\rm bol}$ for the MIR and
FIR cases. We use a constant covering factor $f_{\rm in}=0.39$ to
estimate the torus thickness $H(R_{\rm in})$ at its inner radius.
The slope of the correlation between $M_{\rm bh}-H(R_{\rm in})$ is
$\sim0.23$(see Table 1), which is obviously non-linear. The
correlation between $L_{\rm bol}$ and $H(R_{\rm in})$ has a slope
of $\sim0.37$, which is different from the constant torus
thickness $H$ with luminosity assumed in the receding torus model
suggested by \citet{hgd96}. The further IR observations on a
larger quasar sample including more fainter quasars by the Spitzer
Space Telescope will help understand the physics of the dust tori
in quasars.

\acknowledgments I am grateful to X.Y. Xia for stimulating
discussion, and C. Zier for his helpful explanation of their torus
formation model. I thank Caina Hao for pointing me an error in the
original manuscript. This work is supported by the National
Science Fund for Distinguished Young Scholars (grant 10325314),
NSFC (grants 10173016; 10333020), and the NKBRSF (grant
G1999075403). This research has made use of the NASA/IPAC
Extragalactic Database (NED), which is operated by the Jet
Propulsion Laboratory, California Institute of Technology, under
contract with the National Aeronautic and Space Administration.

\begin{deluxetable}{ccccc}
\tabletypesize{\scriptsize} \tablecaption{Results of statistic
analysis } \tablewidth{0pt} \tablehead{ \colhead{Relation}
&\colhead{Sample } & \colhead{Significant level (per cent) } &
\colhead{$a$} & \colhead{$b$ } } \startdata

$M_{\rm bh}-L_{\rm NIR}/L_{\rm bol}$ & All & 99.93 &
$-0.097\pm0.034$ & $-0.125\pm0.279$ \\
~~ & BL quasars & 99.94 & $-0.149\pm0.038$ & ~~$0.325\pm0.322$ \\
$M_{\rm bh}-L_{\rm MIR}/L_{\rm bol}$ & All & 99.94 &
$-0.135\pm0.040$ & ~~$0.224\pm0.335$ \\
~~ & BL quasars & 99.85 & $-0.169\pm0.045$ & ~~$0.526\pm0.381$ \\
$M_{\rm bh}-L_{\rm FIR}/L_{\rm bol}$ & All & 99.99 &
$-0.239\pm0.058$ & ~~$0.658\pm0.475$ \\
~~ & BL quasars & 99.88 & $-0.255\pm0.073$ & ~~$0.809\pm0.625$ \\
$M_{\rm bh}-L_{\rm IR}/L_{\rm bol}$ & All & 99.98 &
$-0.151\pm0.040$ & ~~$0.697\pm0.330$ \\
~~ & BL quasars & 99.93 & $-0.178\pm0.044$ & ~~$0.941\pm0.376$ \\
$M_{\rm bh}-H_{\rm in}$ & All & 99.51 &
~~$0.232\pm0.054$ & $-2.825\pm0.452$ \\
~~ & BL quasars & 99.51 & ~~$0.275\pm0.061$ & $-3.220\pm0.523$ \\
$L_{\rm bol}-H_{\rm in}$ & All & $>99.99$ &
~~$0.367\pm0.051$ & ~~$-17.760\pm2.320$ \\
~~ & BL quasars & $>99.99$ & ~~$0.344\pm0.044$ & ~~$-16.708\pm2.019$ \\
$M_{\rm bh}-H/R$ & All(50) & $99.93$ &
$-0.152\pm0.045$ & ~~$1.011\pm0.376$ \\
~~ & BL quasars(42) & $99.93$ & $-0.210\pm0.054$ & ~~$1.532\pm0.461$ \\
 \enddata
 \tablecomments{Columns (3) and (4): coefficients of the linear regression:
 $\log_{10}Y
=a\log_{10}X+b$. }

\end{deluxetable}

\end{document}